\documentclass[onecolumn,showpacs,prc,floatfix,12pt]{revtex4}
\usepackage{graphicx}
\usepackage{amsmath}
\newcommand{\p}[1]{(\ref{#1})}
\newcommand\beq{\begin{eqnarray}} \newcommand\eeq{\end{eqnarray}}
\newcommand\beqstar{\begin{eqnarray*}} \newcommand\eeqstar{\end{eqnarray*}}
\newcommand{\beqe}{\begin{equation}} \newcommand{\eeqe}{\end{equation}}
\newcommand{\bal}{\begin{align}}

\begin {document}
\title{Competition of ferromagnetic and antiferromagnetic \protect\\spin ordering   in
nuclear matter }
\author{ A. A. Isayev}
\affiliation{Kharkov Institute of Physics and Technology,
Academicheskaya Str. 1,
 Kharkov, 61108, Ukraine}
 \date{\today}
\begin{abstract}In the framework of a Fermi liquid theory it is
considered the possibility of ferromagnetic and antiferromagnetic
phase transitions in symmetric nuclear matter with Skyrme
effective interaction. The zero temperature dependence of
ferromagnetic and  antiferromagnetic spin polarization parameters
as functions of density  is found for SkM$^*$, SGII effective
forces. It is shown that in the density  domain, where both type
of solutions of self--consistent equations exist, ferromagnetic
spin state is more preferable than antiferromagnetic one.
\end{abstract}
 \maketitle

\section{Introduction}
The appearance of spontaneous spin polarized states in nuclear
matter is the topic of a great current interest due to relevance
in astrophysics. According as nuclear matter is spin polarized or
not, drastically different scenarios of supernova explosion and
cooling of neutron stars can be realized. The possibility of a
phase transition of normal nuclear matter to ferromagnetic state
was studied by many authors. In Ref.~\cite{R} within a hard sphere
gas model  it was shown that neutron matter becomes ferromagnetic
at $\varrho\approx0.41\,\mbox{fm}^{-3}$. In Refs.~\cite{S,O} it
was found that the inclusion of long--range attraction
significantly increases the ferromagnetic transition density
(e.g., up to $\varrho\approx2.3\,\mbox{fm}^{-3}$ in the Brueckner
theory with a simple central potential and hard core only for
singlet spin states~\cite{O}). Calculations of magnetic
susceptibility in Ref.~\cite{VNB} with Skyrme effective forces
showed that a ferromagnetic transition occurs at
$\varrho\approx0.18$--$0.26\,\mbox{fm}^{-3}$. The Fermi liquid
criterion for the ferromagnetic instability in neutron matter with
Skyrme interaction is reached at
$\varrho\approx2$--$4\varrho_0$~\cite{RPLP}, $\varrho_0$ being
nuclear matter saturation density. In Ref.~\cite{ALP} there were
formulated general conditions   on the parameters of
neutron--neutron interaction, which  result in the magnetically
ordered state of neutron matter. The spin correlations in dense
neutron matter within the relativistic Dirac--Hartree--Fock
approach with the effective nucleon--meson Lagrangian were studied
in Ref.~\cite{MNQN}, predicting the ferromagnetic transition at
several times nuclear matter saturation density. The importance of
the Fock exchange term in the relativistic mean--field approach
for occurrence of ferromagnetism in nuclear matter was established
in Ref.~\cite{TT}. The stability of strongly asymmetric nuclear
matter with respect to spin fluctuations was investigated in
Ref.~\cite{KW}, where it was shown that even small proton
admixture favors the ferromagnetic instability of the system. This
conclusion was confirmed also by calculations within the
relativistic Dirac--Hartree--Fock approach to strongly asymmetric
nuclear matter~\cite{BMNQ}.

If to consider the models with realistic nucleon--nucleon (NN)
interaction, the ferromagnetic phase transition seems to be
suppressed up to densities well above
$\varrho_0$~\cite{PGS}--\cite{H}. In particular, no evidence of
ferromagnetic instability has been found in recent studies of
neutron matter~\cite{VPR} and asymmetric nuclear matter~\cite{VB}
within the Brueckner--Hartree--Fock approximation with realistic
Nijmegen II, Reid93 and Nijmegen NSC97e NN interactions. The same
conclusion was obtained in Ref.~\cite{FSS}, where magnetic
susceptibility of neutron matter was calculated with the use of
the Argonne $v_{18}$ two--body potential plus Urbana IX
three--body potential.

Here we will continue the study of spin polarizability of nuclear
matter with the use of  effective NN interaction. As a framework
of consideration, it is chosen a Fermi liquid (FL) description of
nuclear matter~\cite{AKP,AIP}. As a potential of NN interaction,
we use the Skyrme effective interaction, utilized earlier in a
number of contexts for nuclear matter
calculations~\cite{SYK}--\cite{AAI}. Since calculations of
magnetic susceptibility with the Skyrme effective forces show,
that nuclear matter undergoes the ferromagnetic phase transition
at some critical density, it is quite natural step to determine
the density dependence of the ferromagnetic spin polarization
parameter of nuclear matter. Besides, in this study we explore the
possibility of antiferomagnetic phase transition in nuclear
matter, when the spins of protons and neutrons are aligned in the
opposite direction. The antiferromagnetic spin polarization
parameter will be determined
 as a function of density as well. Then we study the
question of thermodynamic stability of ferromagnetic and
antiferromagnetic spin states and clarify which phase is
thermodynamically preferable in the density region, where both
solutions of self--consistent equations  exist simultaneously.

Note that we consider  thermodynamic properties of spin polarized
states in nuclear matter up to the high density region relevant
for astrophysics. Nevertheless, we use a pure nucleonic
description of nuclear matter, although other degrees of freedom,
such as pions, hyperons, kaons,  or quarks could be present at
such high densities.
\section{Basic Equations}
 Normal states of nuclear matter are described
  by the normal distribution function of nucleons $f_{\kappa_1\kappa_2}=\mbox{Tr}\,\varrho
  a^+_{\kappa_2}a_{\kappa_1}$
($\kappa\equiv({\bf{p}},\sigma,\tau)$,  ${\bf p}$ is momentum,
$\sigma(\tau)$ is the projection of spin (isospin) on the third
axis, $\varrho$ is the density matrix of the system).  The energy
of the system is specified as a functional of the distribution
function $f$, $E=E(f)$, and determines the single particle energy
 \begin{eqnarray}
\varepsilon_{\kappa_1\kappa_2}(f)=\frac{\partial E(f)}{\partial
f_{\kappa_2\kappa_1}}. \label{1} \end{eqnarray} The
self-consistent matrix equation for determining the distribution
function $f$ follows from the minimum condition of the
thermodynamic potential \cite{AKP} and is
  \begin{eqnarray}
 f=\left\{\mbox{exp}(Y_0\varepsilon+
Y_4)+1\right\}^{-1}\equiv
\left\{\mbox{exp}(Y_0\xi)+1\right\}^{-1}.\label{2}\end{eqnarray}
Here the quantities $\varepsilon,Y_4$ are matrices in the space of
$\kappa$ variables, with
$Y_{4\kappa_1\kappa_2}=Y_{4\tau_1}\delta_{\kappa_1\kappa_2}$
$(\tau_1=p,n)$, $Y_0=1/T,\ Y_{4p}=-\mu_p^0/T$ and
$Y_{4n}=-\mu_n^0/T$
 the Lagrange multipliers, $\mu_p^0$ and $\mu_n^0$  the chemical
potentials of protons and neutrons and $T$  the temperature. We
shall study the possibility of formation of different types of
spin ordering (ferromagnetic and antiferromagnetic) in nuclear
matter.

The normal  distribution function can be expanded in the Pauli
matrices $\sigma_i$ and $\tau_k$ in spin and isospin
spaces
\begin{align} f({\bf p})&= f_{00}({\bf
p})\sigma_0\tau_0+f_{30}({\bf p})\sigma_3\tau_0\label{7.2}\\
&\quad + f_{03}({\bf p})\sigma_0\tau_3+f_{33}({\bf
p})\sigma_3\tau_3. \nonumber
\end{align}
 For the energy functional, being invariant with
respect to rotations in spin and isospin spaces, the structure of
the single particle energy  is  similar to that of the
distribution function $f$: \begin{align} \varepsilon({\bf p})&=
\varepsilon_{00}({\bf
p})\sigma_0\tau_0+\varepsilon_{30}({\bf p})\sigma_3\tau_0\label{7.3}\\
&\quad + \varepsilon_{03}({\bf
p})\sigma_0\tau_3+\varepsilon_{33}({\bf p})\sigma_3\tau_3.
\nonumber
\end{align}
Using Eqs.~\p{2}, \p{7.3}, it is possible to express evidently the
distribution functions $f_{00},f_{30},f_{03},f_{33}$
 in
terms of the quantities $\varepsilon$: \begin{align}
f_{00}&=\frac{1}{4}\{n(\omega_{+,+})+n(\omega_{+,-})+n(\omega_{-,+})+n(\omega_{-,-})
\},\nonumber
 \\
f_{30}&=\frac{1}{4}\{n(\omega_{+,+})+n(\omega_{+,-})-n(\omega_{-,+})-n(\omega_{-,-})
\},\nonumber\\
f_{03}&=\frac{1}{4}\{n(\omega_{+,+})-n(\omega_{+,-})+n(\omega_{-,+})-n(\omega_{-,-})
\},\nonumber\\
f_{33}&=\frac{1}{4}\{n(\omega_{+,+})-n(\omega_{+,-})-n(\omega_{-,+})+n(\omega_{-,-})
\}.\label{2.4}
 \end{align} Here $n(\omega)=\{\exp(Y_0\omega)+1\}^{-1}$ and
\begin{gather*}
\omega_{+,+}=\xi_{00}+\xi_{30}+\xi_{03}+\xi_{33},\;\\
\omega_{+,-}=\xi_{00}+\xi_{30}-\xi_{03}-\xi_{33},\;\\
\omega_{-,+}=\xi_{00}-\xi_{30}+\xi_{03}-\xi_{33},\;\\
\omega_{-,-}=\xi_{00}-\xi_{30}-\xi_{03}+\xi_{33},\;\end{gather*}
where \begin{align*}\xi_{00}&=\varepsilon_{00}-\mu_{00}^0,\;
\xi_{30}=\varepsilon_{30},\;
\\
\xi_{03}&=\varepsilon_{03}-\mu_{03}^0,\;\xi_{33}=\varepsilon_{33},\\
\mu_{00}^0&={\frac{\mu_p^0+\mu_n^0}{2}},\quad
\mu_{03}^0={\frac{\mu_p^0-\mu_n^0}{2}}.\end{align*}
 As follows from the structure of the distribution
functions $f$, the quantity $\omega_{\pm,\pm}$, being the exponent
in Fermi distribution function $n$, plays the role of the
quasiparticle spectrum. In the considering case the spectrum is
four--fold split due to spin and isospin dependence of the single
particle energy $\varepsilon({\bf p})$ in Eq.~\p{7.3}. The
distribution functions $f$ should satisfy the normalization
conditions \begin{align} \frac{4}{\cal
V}\sum_{\bf p}f_{00}({\bf p})&=\varrho,\label{3.1}\\
\frac{4}{\cal V}\sum_{\bf p}f_{03}({\bf
p})&=\varrho_p-\varrho_n\equiv-\alpha\varrho,\label{3.3}\\
\frac{4}{\cal V}\sum_{\bf p}f_{30}({\bf
p})&=\varrho_\uparrow-\varrho_\downarrow\equiv\Delta\varrho_{\uparrow\uparrow},\label{3.2}\\
\frac{4}{\cal V}\sum_{\bf p}f_{33}({\bf
p})&=(\varrho_{p\uparrow}+\varrho_{n\downarrow})-
(\varrho_{p\downarrow}+\varrho_{n\uparrow})\equiv\Delta\varrho_{\uparrow\downarrow}.\label{3.4}
 \end{align}
 Here $\alpha$ is the isospin asymmetry parameter, $\varrho_{p\uparrow},\varrho_{p\downarrow}$ and
 $\varrho_{n\uparrow},\varrho_{n\downarrow}$ are the proton and
 neutron number densities with spin up and spin down,
 respectively;
 $\varrho_\uparrow=\varrho_{p\uparrow}+\varrho_{n\uparrow}$ and
$\varrho_\downarrow=\varrho_{p\downarrow}+\varrho_{n\downarrow}$
are the nucleon densities with spin up and spin down. The
quantities $\Delta\varrho_{\uparrow\uparrow}$ and
$\Delta\varrho_{\uparrow\downarrow}$ may be regarded as
ferromagnetic (FM) and antiferromagnetic (AFM) spin order
parameters: if all nucleon spins are aligned in one direction
(totally polarized FM spin state), then
$\Delta\varrho_{\uparrow\uparrow}=\varrho$ and
$\Delta\varrho_{\uparrow\downarrow}=0$; if spins of all protons
are aligned in one direction and spins of all neutrons in the
opposite one (totally polarized  AFM spin state), then
$\Delta\varrho_{\uparrow\downarrow}=\varrho$ and
$\Delta\varrho_{\uparrow\uparrow}=0$.

To obtain the self--consistent equations, it is necessary to
specify the energy functional of the system, which we write in the
form
\begin{align} E(f)&=E_0(f)+E_{int}(f), \label{14}\\
{E}_0(f)&=4\sum\limits_{ \bf p}^{} \varepsilon_0({\bf
p})f_{00}({\bf p}),\;\varepsilon_0({\bf p})=\frac{{\bf
p}^{\,2}}{2m_{0}},\nonumber
\\ {E}_{int}(f)&=2\sum\limits_{ \bf p}^{}\{
\tilde\varepsilon_{00}({\bf p})f_{00}({\bf p})+
\tilde\varepsilon_{30}({\bf p})f_{30}({\bf p})\nonumber\\
&\quad+\tilde\varepsilon_{03}({\bf p})f_{03}({\bf p})+
\tilde\varepsilon_{33}({\bf p})f_{33}({\bf p})\} ,
\nonumber\end{align} \begin{align}\tilde\varepsilon_{00}({\bf
p})&=\frac{1}{2\cal V}\sum_{\bf q}U_0({\bf k})f_{00}({\bf
q}),\;{\bf k}=\frac{{\bf p}-{\bf q}}{2}, \nonumber\\
\tilde\varepsilon_{30}({\bf p})&=\frac{1}{2\cal V}\sum_{\bf
q}U_1({\bf k})f_{30}({\bf q}),\nonumber\\ 
\tilde\varepsilon_{03}({\bf p})&=\frac{1}{2\cal V}\sum_{\bf
q}U_2({\bf k})f_{03}({\bf q}), \nonumber\\
\tilde\varepsilon_{33}({\bf p})&=\frac{1}{2\cal V}\sum_{\bf
q}U_3({\bf k})f_{33}({\bf q}). \nonumber
\end{align}
 Here
  $m_0$ is the bare mass of a nucleon, $U_0({\bf k}),...,U_3({\bf k}) $ are the normal FL
amplitudes,
$\tilde\varepsilon_{00},\tilde\varepsilon_{30},\tilde\varepsilon_{03},\tilde\varepsilon_{33}$
are the FL corrections to the free single particle spectrum. With
allowance for Eqs.~\p{1} and \p{14}, we obtain the
self--consistent equations in the form \begin{eqnarray}
\xi_{00}({\bf p})&=&\varepsilon_{0}({\bf
p})+\tilde\varepsilon_{00}({\bf p})-\mu_{00}^0,\;
\xi_{30}({\bf p})=\tilde\varepsilon_{30}({\bf p}), \\
\xi_{03}({\bf p})&=&\tilde\varepsilon_{03}({\bf p})-\mu_{03}^0, \;
\xi_{33}({\bf p})=\tilde\varepsilon_{33}({\bf p}).
\nonumber\end{eqnarray}
  Further for
obtaining numerical results we shall use the Skyrme effective
interaction. In the case of Skyrme forces the normal FL amplitudes
read~\cite{AIP} \begin{align}
U_0({\bf k})&=6t_0+t_3\varrho^\beta
+\frac{2}{\hbar^2}[3t_1+t_2(5+4x_2)]{\bf k}^{2},
\\
U_1({\bf
k})&=-2t_0(1-2x_0)-\frac{1}{3}t_3\varrho^\beta(1-2x_3)\nonumber\\
&\quad-\frac{2}{\hbar^2}[t_1(1-2x_1)-t_2(1+2x_2) ]{\bf
k}^{2}\equiv a+b{\bf k}^{2},
\nonumber\\
U_2({\bf
k})&=-2t_0(1+2x_0)-\frac{1}{3}t_3\varrho^\beta(1+2x_3)\nonumber\\
&\quad-\frac{2}{\hbar^2}[t_1(1+2x_1)- t_2(1+2x_2)]{\bf k}^{2},\nonumber\\
U_3({\bf k})&=-2t_0-\frac{1}{3}t_3\varrho^\beta
-\frac{2}{\hbar^2}(t_1- t_2){\bf k}^{2}\equiv c+d{\bf
k}^{2},\nonumber
\end{align}
where $t_i,x_i,\beta$ are phenomenological constants,
characterizing the given parametrization of Skyrme forces. In
numerical calculations  we
  shall use the  SkM$^*$ \cite{BGH} and  SGII \cite{SG} potentials,
  developed to fit the properties of systems with small isospin
  asymmetry.
 With
account of the evident form of FL amplitudes and
Eqs.~\p{3.1}--\p{3.4}, one can obtain \begin{align}
\xi_{00}&=\frac{p^2}{2m_{00}}-\mu_{00}, \\
\xi_{03}&=\frac{p^2}{2m_{03}}-\mu_{03},\\
\xi_{30}&=(a+b\frac{{\bf
p}^{2}}{4})\frac{\Delta\varrho_{\uparrow\uparrow}}{8}+\frac{b}{32}\langle
{\bf q}^{2}\rangle_{30}, \label{4.3}\\
\xi_{33}&=(c+d\frac{{\bf
p}^{2}}{4})\frac{\Delta\varrho_{\uparrow\downarrow}}{8}+\frac{d}{32}\langle
{\bf q}^{2}\rangle_{33},\label{4.4}
\end{align}
where the effective nucleon mass $m_{00}$ and  effective isovector
mass
  $m_{03}$ are defined by
 the formulae:
\begin{eqnarray}
\frac{\hbar^2}{2m_{00}}&=&\frac{\hbar^2}{2m_0}+\frac{\varrho}{16}
[3t_1+t_2(5+4x_2)],\label{18}\\
\frac{\hbar^2}{2m_{03}}&=& \frac{\alpha\varrho}{16}[t_1(1+2x_1)-
t_2(1+2x_2)],\nonumber\end{eqnarray} and the renormalized chemical
potentials $\mu_{00},\mu_{03}$ should be determined from Eqs.
\p{3.1}, \p{3.3}. In Eqs.~\p{4.3}, \p{4.4} $\langle {\bf
q}^{2}\rangle_{30},\langle {\bf q}^{2}\rangle_{33}$ are the second
order moments of the corresponding distribution functions
\begin{align} \langle {\bf
q}^{2}\rangle_{30}&=\frac{4}{V}\sum_{\bf q}{\bf
q}^2f_{30}({\bf q}),\label{6.1}\\
\langle {\bf q}^{2}\rangle_{33}&=\frac{4}{V}\sum_{\bf q}{\bf
q}^2f_{33}({\bf q}). \label{6.2}\end{align} Thus, with account of
the expressions \p{2.4} for the distribution functions $f$, we
obtain the self--consistent equations \p{3.1}--\p{3.4}, \p{6.1},
\p{6.2} for the effective chemical potentials $\mu_{00},\mu_{03}$,
 FM  and AFM spin
 order parameters
$\Delta\varrho_{\uparrow\uparrow}$,
$\Delta\varrho_{\uparrow\downarrow}$, and the second order moments
$\langle {\bf q}^{2}\rangle_{30},\langle {\bf q}^{2}\rangle_{33}$.
\section{Ferromagnetic and Antiferromagnetic spin order parameters at
zero temperature}

The early researches on spin polarizability  with Skyrme effective
interaction were based on the calculation of magnetic
susceptibility and finding  its pole structure~\cite{VNB,RPLP},
determining the onset of   instability with respect to spin
fluctuations. Here we provide the direct calculation of FM spin
polarization as a function of nuclear matter density at zero
temperature. Besides, we study the possibility of AFM spin
ordering in nuclear matter and competition between these two types
of ordering.

\begin{figure}[p]
\includegraphics[height=12.6cm,width=8.6cm,trim=49mm 105mm 56mm 46mm,
draft=false,clip]{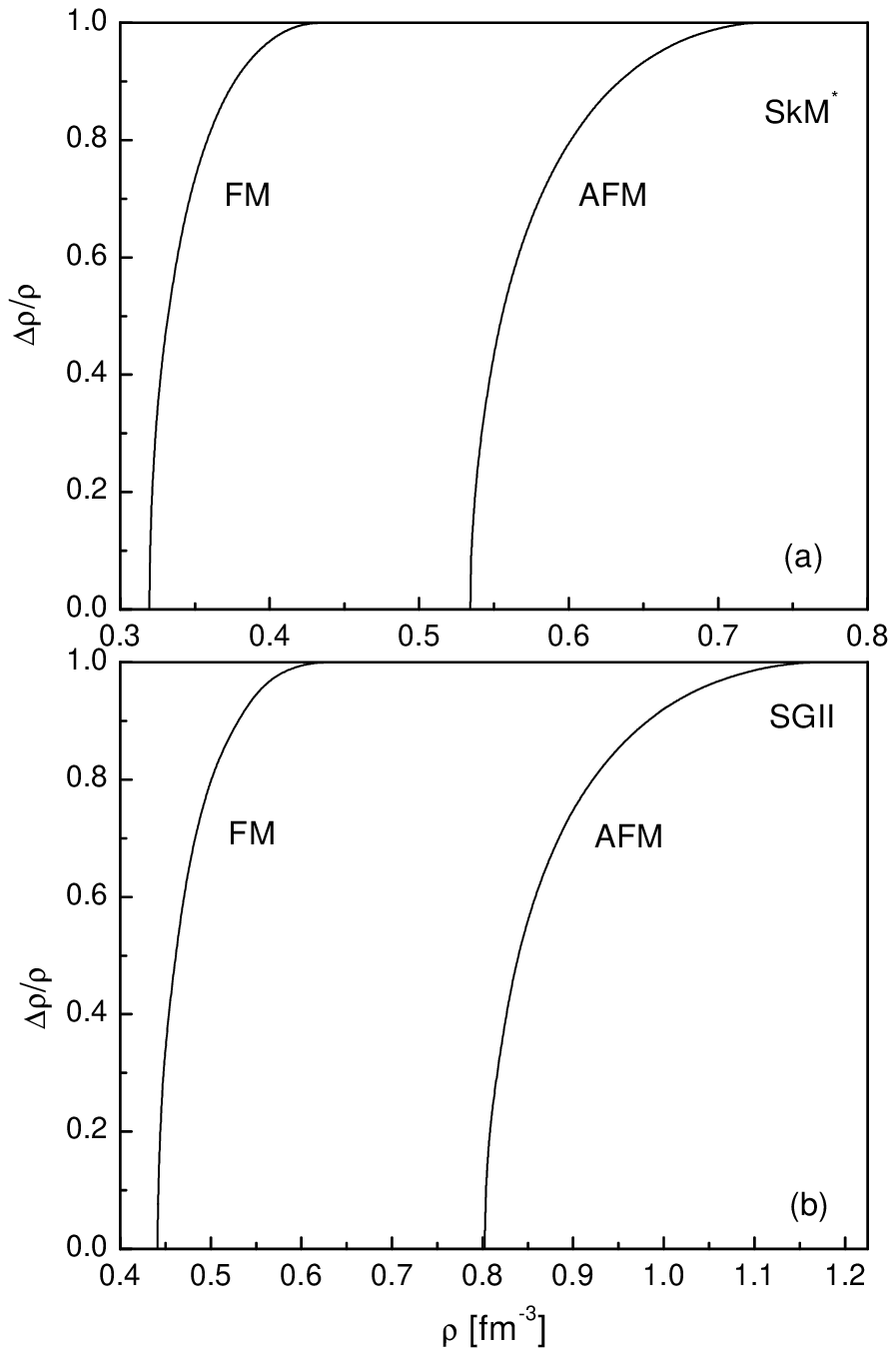} \caption{FM and AFM spin
polarization parameters as functions of density at zero
temperature for (a) SkM$^*$ force and (b) SGII force.
}\label{fig1}
\end{figure}

Let us consider the zero temperature behavior of spin polarization
in symmetric nuclear matter ($\varrho_p=\varrho_n$). The FM spin
ordering corresponds to the case
$\Delta\varrho_{\uparrow\uparrow}\not=0,\langle {\bf
q}^{2}\rangle_{30}\not=0,\Delta\varrho_{\uparrow\downarrow}=0,\langle
{\bf q}^{2}\rangle_{33}=0$, while the AFM spin ordering  to the
case $\Delta\varrho_{\uparrow\downarrow}\not=0,\langle {\bf
q}^{2}\rangle_{33}\not=0,\Delta\varrho_{\uparrow\uparrow}=0,\langle
{\bf q}^{2}\rangle_{30}=0$. In the totally ferromagnetically
polarized state nontrivial solutions of the self--consistent
equations have the form \begin{equation}
\Delta\varrho_{\uparrow\uparrow}=\varrho,\; \langle {\bf
q}^{2}\rangle_{30}=\frac{3}{5}\varrho
k_F^2.\label{18.1}\end{equation} Here $k_F=(3\pi^2\varrho)^{1/3}$
is Fermi momentum  of symmetric nuclear matter in the case when
degrees of freedom, corresponding to spin up of nucleons, are open
while those related to spin down are inaccessible. For totally
antiferromagnetically polarized nuclear matter we have
\begin{equation} \Delta\varrho_{\uparrow\downarrow}=\varrho,\;
\langle {\bf q}^{2}\rangle_{33}=\frac{3}{5}\varrho
k_F^2.\end{equation} The Fermi momentum $k_F$ is given by the same
expression as in Eq.~\p{18.1} since now degrees of freedom,
related to spin down of protons and spin up of neutrons, are
inaccessible.
 The results of numerical
determination of FM $\Delta\varrho_{\uparrow\uparrow}/\varrho$ and
AFM $\Delta\varrho_{\uparrow\downarrow}/\varrho$ spin polarization
parameters are shown in Fig.~\ref{fig1}  for the SkM$^*$ and SGII
effective forces.

The FM spin order parameter arises at density
$\varrho\approx2\varrho_0$ for the SkM$^*$ potential and at
$\varrho\approx2.75\varrho_0$ for the SGII potential. The AFM
order parameter originates at $\varrho\approx3.3\varrho_0$ for the
SkM$^*$ force and at $\varrho\approx5\varrho_0$ for the SGII
force. In both cases FM ordering appears earlier than AFM one.
Nuclear matter becomes totally ferromagnetically polarized
($\Delta\varrho_{\uparrow\uparrow}/\varrho=1$) at density
$\varrho\approx2.7\varrho_0$ for the SkM$^*$ force and at
$\varrho\approx3.9\varrho_0$ for the SGII force. Totally
antiferromagnetically polarized state
($\Delta\varrho_{\uparrow\downarrow}/\varrho=1$) is formed at
$\varrho\approx4.5\varrho_0$ for the SkM$^*$ potential and at
$\varrho\approx7.2\varrho_0$ for the SGII potential.

\begin{figure}[p]
\includegraphics[height=12.6cm,width=8.6cm,trim=49mm 103mm 56mm 46mm,
draft=false,clip]{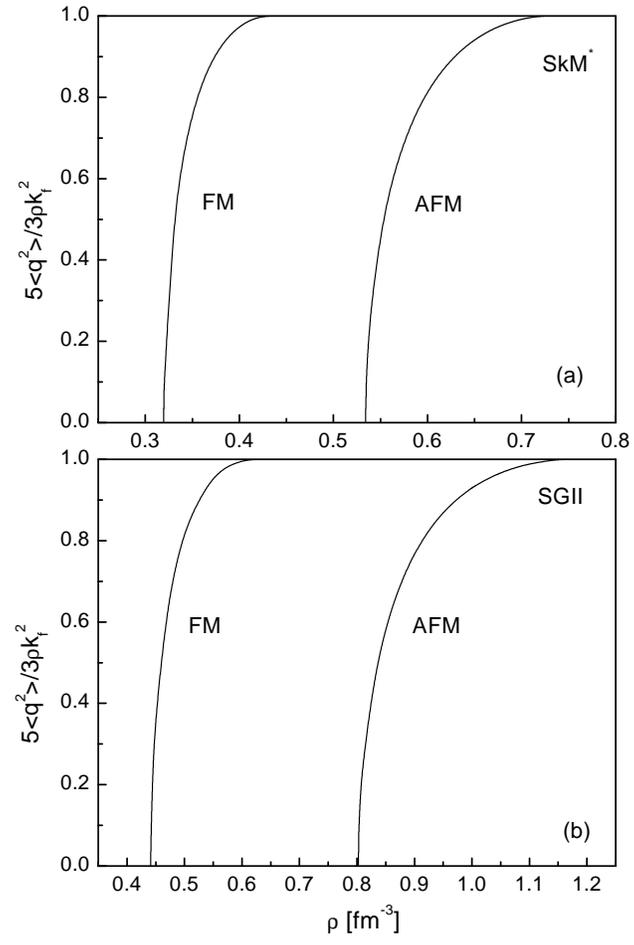} \caption{The same as in
Fig.~\ref{fig1}, but for the second order moments of the
distribution functions. }\label{fig2}
\end{figure}

Note that the second order moments $\langle {\bf
q}^{2}\rangle_{30}, \langle {\bf q}^{2}\rangle_{33}$ of the
distribution functions $f_{30},f_{33}$ play the role of the order
parameters as well. In Fig.~\ref{fig2} it is shown behavior of
these quantities normalized to their  value in totally polarized
state. The ratios $5\langle {\bf q}^{2}\rangle_{30}/3\varrho
k_F^2$ and $5\langle {\bf q}^{2}\rangle_{33}/3\varrho k_F^2$ are
regarded as FM and AFM order parameters, respectively.
The behavior of these quantities is similar to that of the spin
polarization parameters in Fig.~\ref{fig1}, with the same values
of the threshold densities for appearance and saturation of the
order parameters.

\begin{figure}[p]
\includegraphics[height=12.6cm,width=8.6cm,trim=49mm 103mm 56mm 46mm,
draft=false,clip]{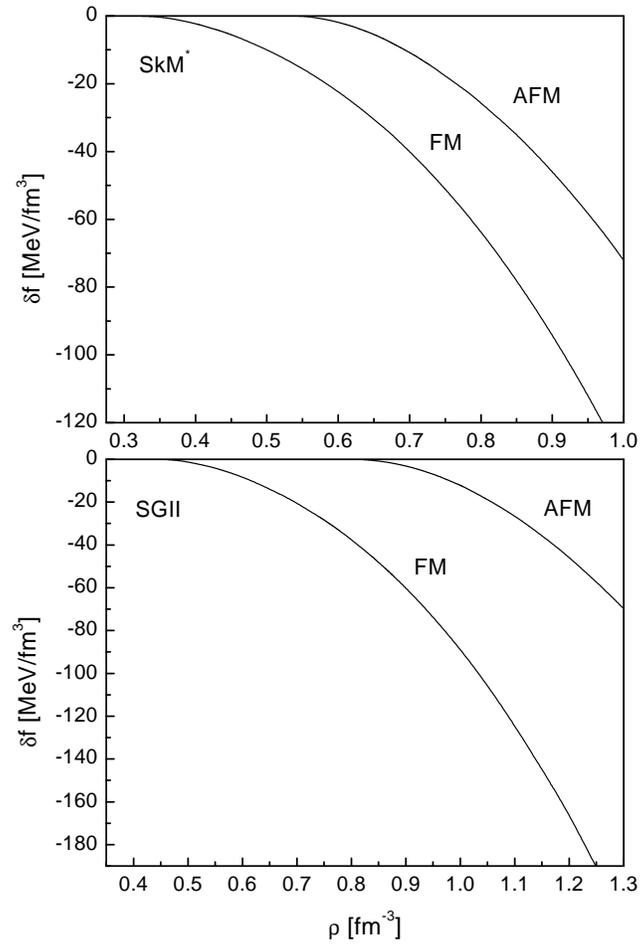} \caption{Free energy density,
measured from that of the normal state, for FM and AFM spin
ordering  as function of density at zero temperature for (a)
SkM$^*$ force and (b) SGII force. }\label{fig3}
\end{figure}

 In the density domain, where FM and AFM solutions of
self--consistent equations exist simultaneously, it is necessary
to clarify, which solution is thermodynamically preferable. With
this purpose it is necessary to compare the free energies of both
states. The results of the numerical calculation of the free
energy density, measured from that of the normal state, are shown
in Fig.~\ref{fig3}.
One can see, that for all relevant densities FM spin ordering is
more preferable than AFM one, and, moreover, the difference
between corresponding free energies becomes larger with increasing
density, so that there is no evidence, that AFM spin ordering
might become preferable at larger densities.

In conclusion, we have considered the possibility of spontaneous
appearance of spin polarized states in symmetric nuclear matter,
corresponding to FM and AFM spin ordering. The study has been done
in the framework of a Fermi liquid description of nuclear matter,
when nucleons interact via Skyrme effective forces (SkM$^*$, SGII
potentials). Unlike the previous considerations, where the
possibility of formation of FM spin polarized states was studied
on the base of calculation of magnetic susceptibility,  we obtain
the self--consistent equations for the FM and AFM spin
polarization parameters and solve them in the case of zero
temperature. It is shown that FM order parameter appears at
densities $2$--$2.75\varrho_0$ and AFM one at densities
$3.3$--$5\varrho_0$. In the density region, where both type of
solutions exist, FM spin ordering wins competition for
thermodynamic  stability.

The author acknowledges the financial support of STCU (grant No.
1480).

\end{document}